\documentclass[sigconf]{acmart}
\AtBeginDocument{%
  }

\copyrightyear{2025}
\acmYear{2025}
\setcopyright{acmlicensed}\acmConference[SIGSPATIAL '25]{The 33rd ACM
International Conference on Advances in Geographic Information
Systems}{November 3--6, 2025}{Minneapolis, MN, USA}
\acmBooktitle{The 33rd ACM International Conference on Advances in
Geographic Information Systems (SIGSPATIAL '25), November 3--6, 2025,
Minneapolis, MN, USA}
\acmDOI{10.1145/3748636.3762779}
\acmISBN{979-8-4007-2086-4/2025/11}

\begin{document}

\title{A Probabilistic Framework for Imputing Genetic Distances in Spatiotemporal Pathogen Models}

\author{Haley Stone}
\affiliation{%
  \institution{University of New South Wales}
  \city{Sydney}
  \country{Australia}}
\email{h.stone@unsw.edu.au}

\author{Jing Du}
\affiliation{%
  \institution{University of New South Wales}
  \city{Sydney}
  \country{Australia}}
\email{jing.du2@unsw.edu.au}

\author{Hao Xue}
\affiliation{%
  \institution{University of New South Wales}
  \city{Sydney}
  \country{Australia}}
\email{hao.xue1@unsw.edu.au	}

\author{Matthew Scotch}
\affiliation{%
  \institution{Arizona State University}
  \city{Phoenix}
  \state{Arizona}
  \country{USA}}
\email{Matthew.Scotch@asu.edu}

\author{David Heslop}
\affiliation{%
  \institution{University of New South Wales}
  \city{Sydney}
  \country{Australia}}
\email{d.heslop@unsw.edu.au	}

\author{Andreas Z\"ufle}
\affiliation{%
  \institution{Emory University}
  \city{Atlanta}
  \state{Georgia}
  \country{USA}}
\email{azufle@emory.edu}

\author{Chandini Raina MacIntyre}
\affiliation{%
  \institution{University of New South Wales}
  \city{Sydney}
  \country{Australia}}
\email{R.MacIntyre@unsw.edu.au}

\author{Flora Salim}
\affiliation{%
  \institution{University of New South Wales}
  \city{Sydney}
  \country{Australia}}
\email{flora.salim@unsw.edu.au}

\renewcommand{\shortauthors}{Stone et al.}

\begin{abstract}
Pathogen genome data offers valuable structure for spatial models, but its utility is limited by incomplete sequencing coverage. We propose a probabilistic framework for inferring genetic distances between unsequenced cases and known sequences within defined transmission chains, using time-aware evolutionary distance modeling. The method estimates pairwise divergence from collection dates and observed genetic distances, enabling biologically plausible imputation grounded in observed divergence patterns, without requiring sequence alignment or known transmission chains. Applied to highly pathogenic avian influenza A/H5 cases in wild birds in the United States, this approach supports scalable, uncertainty-aware augmentation of genomic datasets and enhances the integration of evolutionary information into spatiotemporal modeling workflows.

\end{abstract}



\maketitle

\section{Introduction}

Understanding genetic divergence between pathogen cases is central to many applications in molecular epidemiology, including phylogeographic inference, lineage clustering, and outbreak reconstruction \cite{lemey2009bayesian, grenfell2004unifying}. These divergence values are typically derived from aligned genome sequences using nucleotide substitution models, providing a biologically meaningful measure of relatedness. However, in large-scale surveillance systems, particularly those involving wildlife or decentralized reporting, only a fraction of detected cases are sequenced. This results in sparsely observed divergence matrices that limit the use of genomic structure in downstream analysis. While some modeling frameworks substitute missing divergence values with proxies based on geographic or ecological similarity, such substitutions overlook the evolutionary processes that shape genetic change. Evolutionary divergence is inherently asymmetric and shaped by stochastic mutation processes, host-specific substitution dynamics, and heterogeneous selective pressures. These properties place divergence in a non-Euclidean, tree-like space, where generic similarity measures such as Euclidean distance or latent embeddings can introduce distortion and obscure biological structure \cite{10.1093/sysbio/syr066}. Treating genetic divergence as a standard matrix completion problem may therefore impose structural assumptions misaligned with the geometry of evolutionary processes.

This paper introduces a probabilistic framework for imputing pairwise genetic divergence between pathogen cases using only metadata. The method models conditional quantiles of observed divergence, estimated using substitution models like Kimura's K80, based on features such as spatial distance, temporal lag, host taxonomy, and sampling metadata \cite{kimura1980simple}. By learning directly from real-world divergence patterns, the model provides biologically grounded, uncertainty-aware interval predictions for unsequenced cases.

Our motivating application is the North American spread of highly pathogenic avian influenza A/H5 (clade 2.3.4.4b) between 2021 and 2024, where large-scale wildlife surveillance produced many unsequenced detections \cite{xie2023episodic}. We train a quantile regression model using observed divergence between gene segment sequences and apply it to infer plausible divergence intervals for unsequenced cases. These predictions can support a range of downstream tasks, including lineage-informed spatial clustering, diffusion modeling, genetic anomaly detection, and probabilistic graph augmentation.

Rather than reconstructing phylogenies or estimating explicit transmission chains, our focus is on modeling genetic divergence as a function of spatiotemporal and ecological context. Divergence between cases reflects not only underlying mutation processes, but also the time elapsed and geographic separation between detections. By learning these relationships directly from observed data, the proposed framework enables biologically informed inference in settings where genomic data are sparse but spatial and temporal metadata are reliably available. This allows unsequenced detections to be meaningfully integrated into geospatial models of pathogen spread, supporting downstream applications such as diffusion simulation, contact network construction, and risk-aware clustering in epidemiological graphs. In doing so, our approach bridges molecular evolution and spatial reasoning, offering a principled way to extend genetic signal across incomplete surveillance networks.

Genetic divergence between pathogen cases is a foundational quantity in molecular epidemiology and phylogenetics \cite{lemey2009phylogenetic}. However, estimating divergence without sequence data remains a difficult and underexplored problem. In practical surveillance settings, especially those involving wildlife or decentralized reporting, only a small and often biased fraction of detected cases are genetically sequenced. This results in sparse and uneven divergence matrices that hinder efforts to model transmission, infer genetic clusters, or estimate evolutionary trajectories.

Several factors contribute to the difficulty of this task. First, divergence is defined pairwise, which means that missing values scale quadratically with the number of unsequenced cases. Even modest sequencing gaps can leave most of the matrix unobserved. Second, evolutionary divergence is shaped by complex, asymmetric processes such as mutation accumulation, host-specific substitution rates, and selective pressures. These processes are non-Euclidean and often non-linear, making them poorly suited to standard similarity metrics or latent embedding techniques. Third, there is no ground-truth divergence available for unsequenced case pairs, which rules out direct supervision and complicates model evaluation.

\textbf{This study presents a metadata-driven framework for imputing pairwise genetic divergence using only spatiotemporal and ecological features. The central contribution is a nonparametric quantile regression model that learns conditional divergence intervals from real-world patterns observed in sequenced data.} By predicting full intervals rather than point estimates, the method accounts for uncertainty due to biological variability and incomplete surveillance. The predicted divergence intervals can be used to augment spatial or graph-based models, support lineage-aware clustering, or enable probabilistic diffusion modeling. The approach generalizes beyond avian influenza and offers a scalable, interpretable solution for integrating unsequenced detections into downstream genomic and spatial analyses, including risk mapping, geospatial clustering, and diffusion-based forecasting in sparse-data environments.

Theoretical work in computational phylogenetics has demonstrated that evolutionary divergence cannot be inferred from arbitrary inputs without either sequence data or strong structural assumptions \cite{mossel2005phylogenetics, lemey2009phylogenetic}. While substitution models such as K80, TN93, or GTR provide point estimates of divergence from aligned sequences, they cannot be applied when no alignment is available. Predicting divergence from metadata is therefore fundamentally different from traditional phylogenetic inference: it is a supervised estimation problem with partially observed targets, high-dimensional covariates, and domain-specific structural constraints. To our knowledge, few if any existing models formalize this task explicitly or address its use in graph construction.

The approach proposed in this study addresses this gap by introducing a metadata-driven, supervised model for estimating evolutionary divergence. It learns conditional quantiles of divergence using spatial, temporal, and host-related metadata as predictors, trained on observed pairwise divergence values from sequenced cases. By estimating intervals rather than point predictions, the model accounts for uncertainty in mutation processes and enables probabilistic edge construction in spatial-genetic graphs. This framing supports generalization to unsequenced cases and provides biologically plausible divergence estimates where traditional methods cannot be applied.
Simulation-based studies have shown that bias in input data, such as sampling skew or missing structure, can significantly affect the performance of predictive models in infectious disease settings, especially under spatial or demographic sparsity \cite{zufle2024leveraging}. These findings underscore the need for methods that remain robust when key epidemiological features, including genetic divergence, are incomplete or unevenly observed.

\textbf{Contributions}. Unlike general matrix completion or graph embedding models, the proposed method produces interpretable, metadata-informed outputs that are tailored to the structure of pathogen surveillance data. It avoids alignment, tree inference, and geometric assumptions, instead relying on covariates that are routinely available in outbreak databases. The quantile regression framework enables biologically calibrated uncertainty quantification, which is rarely supported in existing divergence imputation or graph learning methods. By integrating predictive modeling with evolutionary principles, this approach enables graph-based inference in settings with structured genomic sparsity and supports scalable integration of genetic signal into spatiotemporal epidemic models.

All code used in this study is available at \url{https://github.com/cruiseresearchgroup/qri-genetic-distance.git}.

\section{Related Works}

\subsection{Evolutionary distance computation and matching}
Pairwise evolutionary distances form the basis for a range of downstream analyses, including phylogenetic reconstruction, transmission chain inference, lineage tracking, and clustering of infections across space and time \cite{lemey2009bayesian, grenfell2004unifying}. Divergence is commonly computed from aligned genome sequences using nucleotide substitution models, providing a biologically interpretable measure of relatedness between samples. However, in many real-world surveillance systems, genome sequencing is incomplete or unevenly distributed. This results in partially observed divergence matrices that limit not only phylogenetic analysis but also broader tasks such as spatiotemporal modeling, contact tracing, and network-based diffusion estimation. The ability to impute divergence directly from metadata, such as host species, sampling date, and geographic location, offers a way to augment these datasets in a scalable and biologically informed manner, independent of downstream application.

Zoonotic and decentralized reporting systems exhibit highly variable sequencing practices. Surveillance tends to prioritize highly pathogenic cases, known reservoirs, or specific host species, which introduces systematic biases in the observed data \cite{doi:10.1126/science.abo1232}. The resulting evolutionary distance matrices are not only incomplete but structurally sparse, with many cases lacking sequence data entirely. Since genetic distances are pairwise quantities, even moderate rates of missing values rapidly compound, limiting the applicability of graph-based models that rely on sequence-informed edge weights. Models that do not account for missing genetic information often default to suboptimal proxies, such as geographic proximity, which may not accurately reflect transmission potential or evolutionary dynamics.

\subsection{Phylogenetics, genomics and spatial imputation}
Several approaches in spatial computing have been developed to infer missing graph edges or distances. Matrix completion techniques, including low-rank approximation \cite{cai2010singular}, spectral methods \cite{you2020handling}, and graph-based message passing algorithms \cite{grover2019graphite}, aim to estimate unobserved pairwise relationships using geometric or topological assumptions. These approaches often assume symmetric, Euclidean, or latent space structures where distances are smoothly varying and globally constrained. While effective in domains such as recommender systems and network inference, they do not reflect the evolutionary processes governing viral mutation. Genetic divergence is not symmetric, often non-metric, and shaped by factors such as host-specific substitution rates, bottleneck effects, and variable evolutionary pressures over time. As such, general-purpose graph inference methods are poorly suited for modeling missing evolutionary distances in pathogen datasets.

Within genomics, imputation is most commonly associated with reconstructing incomplete sequence data. Genotype imputation techniques rely on population haplotype panels and local linkage disequilibrium to fill in untyped loci, particularly in human genome-wide association studies \cite{marchini2007genotype, das2016next}. These approaches are effective at recovering single-nucleotide polymorphisms (SNPs) when dense reference panels are available, but they focus on within-sequence variation rather than between-sample divergence. Phylogenetic placement tools such as UShER \cite{10.1093/bioinformatics/btr320} and pplacer \cite{darriba2016prediction} allow for inserting partially sequenced samples into pre-existing phylogenies, typically using maximum likelihood or parsimony. However, they require partial sequence data as input, and they do not return explicit divergence estimates usable in continuous or graph-based models. Neither genotype imputation nor placement methods address the problem of estimating divergence between fully unsequenced cases.

Beyond classical tree construction, several probabilistic models have been proposed to estimate evolutionary relationships. Diffusion tree models and coalescent-based inference frameworks treat divergence as a latent variable, estimated from sequence data under assumptions about population dynamics and mutation rates. These models, while theoretically grounded, are typically unsupervised and computationally expensive \cite{mossel2005phylogenetics, lemey2009phylogenetic}. They do not scale to settings with high-dimensional metadata or large numbers of unsequenced cases, and they rarely accommodate auxiliary features such as host identity or geographic metadata.

Recent work has shown that ecological metadata may contain relevant signals for modeling divergence. Host taxonomy, geographic origin, and sampling time have been used to stratify substitution models and to explain rate heterogeneity across lineages \cite{sokurenko2006source, xie2023episodic}. These approaches are valuable for interpreting observed phylogenies, but they are not designed for predictive modeling. Most are retrospective and tree-based, focusing on rate variation rather than on estimating pairwise genetic distances between cases with no sequence data. The integration of such metadata into divergence prediction models remains limited.

Supervised modeling of divergence is less common, despite the availability of large datasets containing both metadata and observed evolutionary distances. Most divergence-aware models are unsupervised, constructed through alignment or tree inference. When machine learning is applied, it is usually for binary classification (e.g., same lineage vs. different lineage) or for learning low-dimensional embeddings from sequences, not for predicting continuous divergence values. One notable challenge is that divergence is a noisy, non-i.i.d. target that depends on underlying phylogenetic constraints, time since infection, and molecular substitution mechanisms. These characteristics violate common assumptions in regression modeling, and necessitate methods that accommodate non-linearity, heteroscedasticity, and biological interpretability.

\subsection{Deep generative models}
In recent years, deep generative models have been applied to missing genetic data reconstruction. Variational autoencoders (VAEs) and transformer-based language models have been used to learn latent representations of aligned sequences and infer plausible reconstructions \cite{li2024discdiff, yelmen2023overview, druet2014toward}. However, these methods are typically trained on dense genomic alignments, assume uniform sampling, and focus on generating full sequences or masked regions rather than estimating inter-sample divergence. They also rarely integrate spatiotemporal metadata or epidemiological features and are not designed to support large-scale inference under surveillance-driven sparsity. Moreover, these models typically lack uncertainty quantification and interpretability, limiting their utility for decision support in public health or field-based applications.

\begin{figure*}
    \centering
    \includegraphics[width=0.8\linewidth]{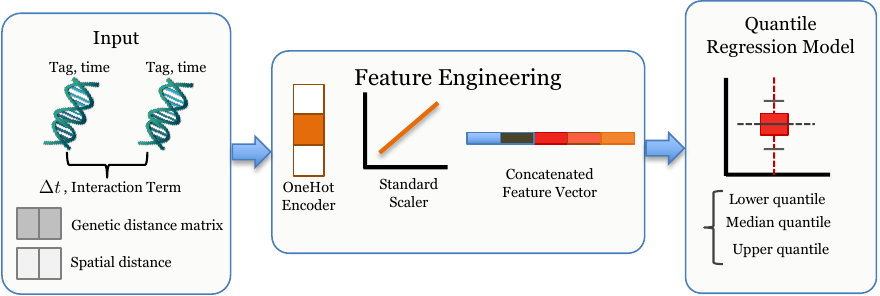}
    \caption{Proposed quantile modeling pipeline for genetic divergence imputation}
    \label{fig:pipeline}
    \vspace{-5mm}
\end{figure*}

\section{Preliminaries}

\subsection{Overview}

This study introduces a spatially-aware modeling framework for estimating evolutionary divergence between avian influenza A/H5 case pairs using pairwise spatiotemporal and host metadata. The objective is to generate plausible divergence intervals for unsequenced cases, enabling structured evolutionary inference even in the presence of sparse genomic data. Furthermore, the model does not infer transmission paths or reconstruct full sequences, but rather estimates plausible divergence values grounded in real-world sequence data, supporting integration of unsequenced detections into distance-aware models.

Each pairwise input is defined by features such as temporal lag, geographic distance, and host taxonomy, which jointly reflect heterogeneity in viral transmission and mutation processes. The model estimates conditional quantiles of genetic divergence, providing interval-valued outputs that capture uncertainty in evolutionary rates due to biological variation and sampling gaps. By learning from real-world divergence patterns across observed sequences, the model enables extrapolation to unsequenced contexts, facilitating comparison of evolutionary patterns across space, time, and host ecology. The framework supports multiple downstream applications, including spatial heterogeneity analyses, detection of anomalous divergence, and imputation in incomplete evolutionary datasets.

\subsection{Evolutionary Distance Computation}

The central modeling target in this study is genetic divergence between pairs of avian influenza A/H5 virus sequences. Unlike structural similarity measures common in computer science, such as Levenshtein or Longest Common Subsequence (LCSS), which compute edit distances based on character alignment or string operations, evolutionary models are designed to represent the probabilistic accumulation of molecular changes over time. These models embed domain-specific assumptions, including nucleotide substitution biases and unobserved transitions at the same site, which are needed for interpreting genetic relatedness in a biologically meaningful way.

In this study, genetic divergence is estimated using the Kimura two-parameter (K80) model \cite{kimura1980simple}. The K80 model distinguishes between transitions and transversions, assigning separate rates for these two classes of substitution. This distinction reflects observed substitution asymmetries in RNA viruses and provides improved accuracy over simpler models like Jukes-Cantor (JC69), which assume equal substitution rates across all nucleotide pairs. More complex models such as Tamura-Nei (TN93) allow for unequal base frequencies and differential transition rates, but were not used due to their increased parameterisation and marginal performance gain within a single clade.

For two aligned sequences of length \( L \), let \( P \) denote the proportion of sites with transitions and \( Q \) the proportion with transversions. The K80 distance \( d \) is defined as:

\[
d = -\frac{1}{2} \ln(1 - 2P - Q) - \frac{1}{4} \ln(1 - 2Q)
\]

This formulation corrects for the underestimation of divergence that arises from multiple unobserved substitutions at the same nucleotide position. Because it operates under assumptions appropriate for within-clade comparisons, the K80 model yields real-valued distances that retain both phylogenetic meaning and numerical stability, allowing integration into subsequent learning algorithms. The resulting pairwise values are treated as the target variable in the quantile regression framework described below.

\subsection{Sequence Matching and Dataset Construction}

All evolutionary distances used in this study are derived from the Avian-US dataset introduced by Du et al. (2025) \cite{du2025blue}, a curated record of highly pathogenic avian influenza (HPAI) A/H5 wild bird detections in the United States between 2021 and 2024. Because metadata for sequenced isolates are manually entered and often inconsistently structured, a taxonomy-aware record linkage model was used to match viral sequences to associated case records. The matching framework incorporated approximate geographic alignment, temporal proximity, and fuzzy taxonomic similarity based on standard avian classification systems.

A supervised model was developed to classify record pairs as matches or non-matches using features that capture spatial, taxonomic, and temporal relationships. The model used an XGBoost classifier trained on a manually labeled dataset of positive and negative matches, with parameters tuned for high precision under class imbalance. Only confidently matched records were retained for downstream analysis. We evaluated match precision on a held-out manually reviewed subset, achieving >95\% precision, ensuring that the training set reflects ground truth divergence pairs. Sequences originating from poultry or with ambiguous host labels were excluded to maintain ecological consistency within the wild bird subset.

Following preprocessing and record linkage, we identified 2,143 unique wild bird detections with complete metadata and high-quality genetic sequences. These yielded 1,048,575 valid sequence same year pairs with computable evolutionary distances. Each pair was annotated with spatial, temporal, and host features derived from case metadata. To support supervised learning, the dataset was structured in pairwise long form, where each row corresponds to a unique case pair with its associated feature vector and genetic divergence. This representation offers three key advantages: (i) it eliminates the redundancy and memory overhead of square distance matrices, (ii) it enables direct encoding of asymmetric pairwise metadata into feature vectors, and (iii) it permits efficient exclusion of incomplete or biologically implausible pairs without requiring matrix reconstruction. The resulting long-form dataset is specifically tailored for integration into regression-based models, allowing flexible and scalable estimation of divergence across heterogeneous metadata-defined contexts.

This format allows the framework to operate directly on pairwise metadata, enabling flexible modeling of evolutionary divergence using observable covariates. The cleaned, matched, and annotated pairwise dataset constitutes both the empirical substrate and structural basis for the quantile regression framework described in the next section.

\section{Methodology}
\subsection{Problem Formulation}

Let \( S = \{s_1, s_2, \dots, s_n\} \) be the set of avian influenza A/H5 cases with available  gene segment sequences, and let \( A = \{a_1, \dots, a_m\} \) be the set of cases without sequence data. For each ordered pair \( (i, j) \), where \( i \in S \cup A \) and \( j \in S \), define a metadata feature vector \( x_{ij} \in \mathbb{R}^p \), constructed from spatial, temporal, and host-specific covariates. The observed genetic divergence \( d_{ij} \) is computed using the Kimura two-parameter (K80) distance for all pairs \( (i, j) \in S \times S \).

The objective is to estimate the conditional quantile function \( Q_{\tau}(d_{ij} \mid x_{ij}) \) for \( \tau \in \{0.05, 0.5, 0.95\} \), corresponding to the lower, median, and upper bounds of predicted genetic divergence. This formulation yields interval-valued predictions that reflect biological and sampling uncertainty in mutation accumulation.

The feature space \( x_{ij} \) includes both symmetric quantities, such as spatial and temporal separation, and asymmetric metadata, such as host family identity and sample year. These features are constructed to preserve directional relationships required for inference involving unsequenced cases. Model training is restricted to the subset \( S \times S \), where divergence values are observed. At inference time, the trained model is applied to the set \( A \times S \), where \( d_{ij} \) is unobserved but covariates are available.

This formulation enables predictive inference of evolutionary divergence for unsequenced detections using metadata alone. The resulting divergence intervals can be incorporated into spatial diffusion models or graph-based epidemiological frameworks that require genetic distance as an edge attribute.

\subsection{Feature Construction}

Each feature vector \( x_{ij} \) is constructed from metadata associated with case \( i \) and reference case \( j \). Temporal separation is encoded as the absolute difference in collection dates, measured in days: \( \Delta t_{ij} = |t_i - t_j| \). Spatial distance is computed using the great-circle distance (in kilometers) between the two samples were collected. Let \( d_{\text{geo},ij} \) denote this value.

Host-related features are derived from the taxonomic metadata. Each host species is assigned a family based on avian taxonomic classification, and family identity is encoded using one-hot vectors. To capture potential phylogenetic or ecological constraints on mutation rates, a binary indicator variable \( \texttt{same\_family}_{ij} \) is included, equal to 1 if both cases belong to the same avian family and 0 otherwise. A second binary indicator \( \texttt{same\_state}_{ij} \) encodes whether both cases were reported in the same state.

A spatial-temporal interaction term \( \Delta t_{ij} \cdot d_{\text{geo},ij} \) is included to capture the effect of spatiotemporal coupling on divergence, allowing for potential acceleration or attenuation of mutation accumulation depending on the dynamics of spread. This interaction captures acceleration or attenuation of divergence under rapid diffusion or geographically restricted outbreaks. Additionally, the sample year is treated as a categorical variable to account for differences in surveillance intensity and lineage prevalence across time.

Before model training, all continuous variables are standardised to zero mean and unit variance. Categorical variables are expanded via one-hot encoding to enable compatibility with tree-based estimators. These transformations standardize the feature space and enable robust learning of complex interactions among epidemiologically relevant covariates.In addition to separate temporal and geographic predictors, we explicitly model their joint effect via the interaction term $\Delta t_{ij} \cdot d_{\mathrm{geo},ij}$ in order to capture spatiotemporal coupling, where rapid geographic spread over short intervals or prolonged temporal gaps at short distances may alter expected divergence.

For interpretability, we computed SHAP values for the median quantile model ($\tau{=}0.5$) using a tree-based explainer on the held-out test folds. We averaged absolute SHAP values per feature within each divergence stratum (LOW/MID/HIGH) and normalized them so importances sum to $100\%$ per stratum. The resulting shares are summarized in Fig.~\ref{fig:shap-heatmap}; per-feature values and scripts are provided in the repository.

\subsection{Quantile Regression}

To estimate plausible intervals of genetic divergence given metadata features, we model the conditional quantile function $Q_{\tau}(d_{ij} \mid x_{ij})$ for divergence $d_{ij}$ between case pairs $i$ and $j$ at quantile level $\tau$. Formally, given a random variable $d_{ij}$ with conditional cumulative distribution function $F(d_{ij} \mid x_{ij})$, the $\tau$-th conditional quantile function is defined as:

\begin{equation}
Q_{\tau}(d_{ij} \mid x_{ij}) = \inf \left\{ q \in \mathbb{R} \mid F(d_{ij} \leq q \mid x_{ij}) \geq \tau \right\}
\end{equation}

To estimate this conditional quantile function from data, we minimize the quantile loss (also known as pinball loss) given by:

\begin{equation}
L_\tau(y, \hat{y}) = \sum_{i=1}^N \rho_\tau(y_i - \hat{y}_i), \quad \text{where} \quad \rho_\tau(u) = u \cdot (\tau - \mathbb{I}(u < 0))
\end{equation}

where $y_i$ represents the observed K80 divergence for a given case pair $i$, $\hat{y}_i$ is the predicted quantile, and $\mathbb{I}(\cdot)$ is the indicator function. 

The pinball loss assigns asymmetric penalties to over- and under-estimation, proportional to the selected quantile $\tau$. We chose quantiles $\tau$ =0.05,0.5,0.95 to capture high-confidence intervals suitable for use in probabilistic edge construction within diffusion graphs. 

Quantile regression is particularly suited for modeling skewed biological distributions because it estimates multiple points (quantiles) of the conditional distribution, capturing heteroscedasticity and non-symmetrical behavior of residuals commonly observed in evolutionary data. Gradient boosting methods iteratively minimize prediction error through an ensemble of weak learners, effectively modeling non-linear relationships and interaction effects among spatial, temporal, and host features.

Mathematically, the boosting procedure for quantile regression solves:

\begin{equation}
f_m(x) = f_{m-1}(x) + \arg\min_{h} \sum_{i=1}^{N} \rho_{\tau} \left( y_i - \left( f_{m-1}(x_i) + h(x_i) \right) \right)
\end{equation}

where \( f_m(x) \) is the model prediction at iteration \( m \), \( \rho_{\tau}(u) = u(\tau - \mathbb{I}(u < 0)) \) is the quantile loss function, and \( h(x) \) represents the incrementally added weak learner. This framework enables flexible approximation of conditional quantiles even in the presence of complex, high-dimensional covariate structures.

Gradient-boosted decision trees are well suited to quantile regression due to their ability to model non-linear interactions. We implement this using the \texttt{lightgbm} framework, fitting separate models for quantiles $\tau \in \{0.05, 0.5, 0.95\}$. This setup produces predictive intervals that represent lower bound, median, and upper bound of divergence, supporting interval-based inference.

\section{Experiments}
\subsection{Model Training and Application}

The model is trained on case pairs with observed K80 distances. To improve generalization, the training data are stratified by geographic region and sample year. 

Spatial autocorrelation, where observations close in space and time exhibit correlated residuals, can violate the independence assumption essential for unbiased estimation. To mitigate this, we perform stratification by geographic region and year, then subsample randomly within each stratum. This stratified subsampling reduces spatial clustering and ensures broader representativeness, which mathematically enhances generalization by preserving variance structure across training folds.

Continuous features such as temporal lag and geographic distance are standardized as:

\begin{equation}
x' = \frac{x - \mu\_x}{\sigma\_x}
\end{equation}

where $\mu_x$ and $\sigma_x$ are the mean and standard deviation of the feature $x$ across the training set. Categorical variables are one-hot encoded.

Hyperparameters, including tree depth, learning rate, and number of estimators, are selected by grid search with five-fold cross-validation. The objective is to minimize mean quantile loss:

\begin{equation}
\min\_{\theta} \frac{1}{K} \sum\_{k=1}^{K} L\_{\tau}\left(y^{(k)}, f\_{\theta}(X^{(k)})\right)
\end{equation}

At inference, the model is applied to each unsequenced case $a \in A$ paired with each reference case $s_j \in S$. The model outputs $[Q_{0.05}(x_{aj}), Q_{0.95}(x_{aj})]$, quantifying the predicted divergence range.

Validation is conducted on a held-out set disjoint in space and time from training. Coverage probability is computed as:

\begin{equation}
C = \frac{1}{N_{\text{test}}} \sum_{i=1}^{N_{\text{test}}} \mathbb{I}\left( y_i \in \left[ Q_{0.05}(x_i),\ Q_{0.95}(x_i) \right] \right)
\end{equation}

\subsection{Divergence Range Stratification}

To assess generalizability under varying divergence conditions, case pairs were stratified into three K80 divergence classes based on quantiles of the observed distribution (Table \ref{tab:k80-strata}). This captures differences in evolutionary signal, from near-identical outbreak samples to inter-lineage comparisons.

\begin{table}[H]
\centering
\caption{K80 divergence strata used for evaluation}
\label{tab:k80-strata}
\begin{tabular}{lcc}
\toprule
\textbf{Class} & \textbf{Distance Range} & \textbf{Mean Distance} \\
\midrule
LOW  & [0.000000, 0.003246]     & $\approx$ 0.0020 \\
MID  & [0.003246, 0.006632]     & $\approx$ 0.0045 \\
HIGH & [0.006632, 0.492703]     & $\approx$ 0.0193 \\
\bottomrule
\end{tabular}
\end{table}

Stratified 5-fold cross-validation was performed by region and year. For each test fold, predicted 5th, 50th, and 95th quantiles were computed. Evaluation was based on residual error and 90\% interval coverage.

\subsection{Ablation Analysis}
To evaluate the contribution of individual covariates to divergence estimation, we conducted a series of ablation experiments. In each experiment, one feature was removed from the input set, the model was retrained from scratch, and its predictions were compared to the full-feature baseline. 

The primary metric was the change in average predicted interval width, defined as:
\begin{equation}
\Delta W = \bar{W}_{\text{ablated}} - \bar{W}_{\text{full}}
\end{equation}
where $\bar{W}$ denotes the mean width of the 90\% prediction interval across all test pairs. A positive $\Delta W$ indicates that the removed feature contributed to narrower, more confident interval estimates.

In addition to $\Delta W$, we recorded shifts in median absolute error (MAE) and interval coverage to assess the predictive and calibration impact of each feature. The resulting differences across ablation runs quantify the marginal influence of each covariate on interval width, point prediction accuracy, and calibration stability.

\subsection{Downstream Integration}

The imputed divergence intervals are intended for integration into a heterogeneous graph framework for spatiotemporal forecasting of avian influenza, specifically the BLUE model proposed by Du et al. (2025) \cite{du2025blue}. In this model, each node corresponds to a reported case, and edges represent epidemiological relationships derived from spatial proximity, host ecology, environmental covariates, and genetic similarity.

Genetic similarity is encoded as edge weights based on observed evolutionary distances between sequences. However, for unsequenced cases, no such edge can be constructed directly. The proposed quantile regression model addresses this limitation by estimating biologically plausible divergence intervals for each unsequenced case relative to the sequenced reference set. The midpoint or lower bound of the predicted interval can be used as a proxy distance, while the full interval allows for uncertainty-aware edge construction or probabilistic graph augmentation.

Once imputed, these genetic distances enable the formation of additional edges between unsequenced and sequenced nodes in the graph, allowing partially observed cases to contribute to diffusion modeling, clustering, or downstream predictive tasks. This preserves the structural integrity of the evolving epidemic network and improves the ability of the model to forecast transmission risk in settings where genomic data are sparse or incomplete.

\section{Results}
\subsection{Overall Performance}
We observed high-level performance of the quantile regression imputation (QRI) model across the three genetic divergence strata: LOW, MID, and HIGH. Table 1 presents median prediction accuracy (50\% quantile) and coverage of the 90\% prediction interval. 

\begin{table}[H]
\centering
\caption{QRI median prediction accuracy and 90\% coverage by divergence class}
\label{tab:qri-performance}
\begin{tabular}{lcccc}
\toprule
\textbf{Class} & \textbf{MAE} & \textbf{RMSE} & \textbf{$R^2$} & \textbf{Coverage (\%)} \\
\midrule
LOW & 0.00060 & 0.00077 & 0.03 & 94.95 \\
MID & 0.00064 & 0.00080 & 0.07 & 92.63 \\
HIGH & 0.01074 & 0.04352 & 0.03 & 90.07 \\
TOTAL & 0.00283 & 0.00542 & 0.19 & 92.6 \\
\bottomrule
\end{tabular}
\end{table}

The QRI model achieved low prediction error for LOW and MID divergence pairs (MAE < 0.001), and maintained calibrated 90 percent prediction intervals across all divergence strata. While RMSE and $R^2$ were lower in the HIGH class, reflecting the reduced informativeness of spatial and temporal covariates at greater evolutionary distances, the model still produced well-calibrated uncertainty bounds. This reliability is important for reconstructing genetic-spatial graphs in surveillance settings where many cases lack sequence data. By enabling edge imputation between unsequenced nodes, the model supports spatial inference tasks such as diffusion simulation, phylogeographic clustering, and contact network augmentation.

Although $R^2$ values appear low within each divergence class, this partly reflects the evaluation setup. The baseline Mean Model used in these calculations has access to the true class label and computes variance relative to the class-specific mean. However, our model makes predictions without that information, relying solely on the available metadata. When performance is assessed across the full dataset without stratification, the model achieves $R^2$ of 0.19. 
    
To understand this drop in performance in the HIGH divergence class, we examined the underlying spatiotemporal structure of the data. HIGH divergence pairs were, on average, collected further apart in both time and space. These high-divergence cases also showed increased host diversity, consistent with inter-clade or cross-species comparisons. The median temporal gap increased from 66 days in the LOW class to 93 days in the HIGH class, while the median spatial distance rose from 1275 km to 2349 km. This confirms that increased divergence tends to coincide with broader temporal and geographic separation, which likely reduces the discriminative value of the covariates used in the model. The QRI framework remains most effective for short- to mid-range divergence where spatiotemporal metadata retains explanatory power.

\subsection{Quantile Interval Calibration}

To assess the calibration of predicted uncertainty, we compared the real-world coverage of the 90\% prediction intervals across the LOW, MID, and HIGH divergence strata. The model achieved strong alignment with the nominal target: real-world coverage was 94.9\% in the LOW class, 92.6\% in MID, and 90.1\% in HIGH.

Interval widths increased substantially across strata, reflecting the model’s sensitivity to evolutionary uncertainty. The average width rose from 0.0022 in the LOW class to 0.0027 in MID and reached 0.0605 in HIGH, an approximate 30-fold increase. These wider intervals in the HIGH divergence group reflect the diminished explanatory power of spatial and temporal covariates when ecological heterogeneity and lineage diversity are greater.

Despite this variance, the model maintained sharp and well-calibrated intervals in LOW and MID strata, with mean absolute errors (MAE) of 0.00061 and 0.00064, respectively. In the HIGH class, the MAE increased to 0.0107, consistent with broader uncertainty in long-distance, cross-host scenarios. 

\subsection{Residual Distribution by Divergence Class}

\begin{figure}[H]
  \centering
  \includegraphics[width=\linewidth]{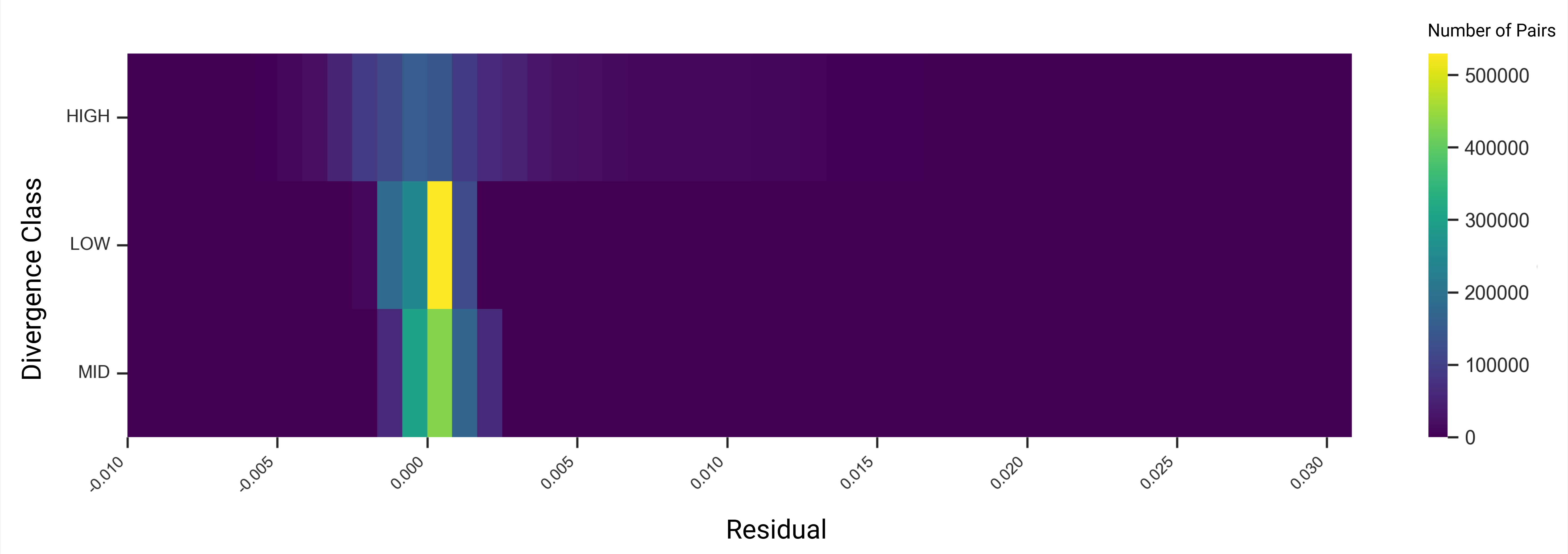}
  \caption{Residual density heatmap by divergence class. Residuals are binned and counts aggregated per class.}
  \label{fig:residual-heatmap}
  \vspace{-5mm}
\end{figure}

To further examine the distribution of prediction residuals across divergence strata, we constructed a heatmap of residual density by class (Figure~\ref{fig:residual-heatmap}). Residuals were binned and aggregated to visualize the concentration of prediction errors. As expected, the LOW and MID divergence classes show sharp concentration near zero, indicating well-calibrated predictions with narrow uncertainty. In contrast, the HIGH class exhibits broader dispersion across bins, consistent with increased uncertainty and less predictable divergence. This visualization supports the trend observed in the error metrics, highlighting a clear stratification in prediction confidence across evolutionary distance classes.

\subsection{Ablation Study}
We conducted a covariate ablation study to evaluate the relative contribution of each feature to the QRI model. For each divergence class, we retrained the model after removing one covariate at a time and computed the change in median prediction error (MAE) and average interval width (95\% - 5\%).

\begin{table}[H]
\centering
\caption{Effect of covariate removal on prediction error and interval width across all divergence classes}
\label{tab:ablation}
\begin{tabular}{lcc}
\toprule
\textbf{Feature Removed} & \textbf{$\Delta$ MAE} & \textbf{$\Delta$ Interval Width} \\
\midrule
$\Delta t$ & +0.00021 & +0.00085 \\
Physical Distance & +0.00013 & +0.00049 \\
Same Taxonomic Family & +0.00009 & +0.00026 \\
Interaction Term & +0.00015 & +0.00037 \\
Year & +0.00007 & +0.00031 \\
\bottomrule
\end{tabular}
\end{table}

Temporal and spatial covariates, particularly  $\Delta$ t and Physical Distance, contributed most to predictive accuracy and interval sharpness. Removing categorical features such as taxonomic family had lesser but non-negligible effects, supporting the inclusion for host-specific modeling. Feature contributions were not strictly additive. Spatial and temporal variables had compounded effects when removed together, resulting in greater degradation than either alone. This reflects the model's ability to exploit joint spatiotemporal structure to constrain divergence estimates. Categorical metadata, such as taxonomic family, contributed lesson their own but provided meaningful constrains in settings where primary spatiotemporal signals were weak or ambiguous. 

\subsection{Feature Importance via SHAP}
\begin{figure}[H]
\centering
\includegraphics[width=\linewidth]{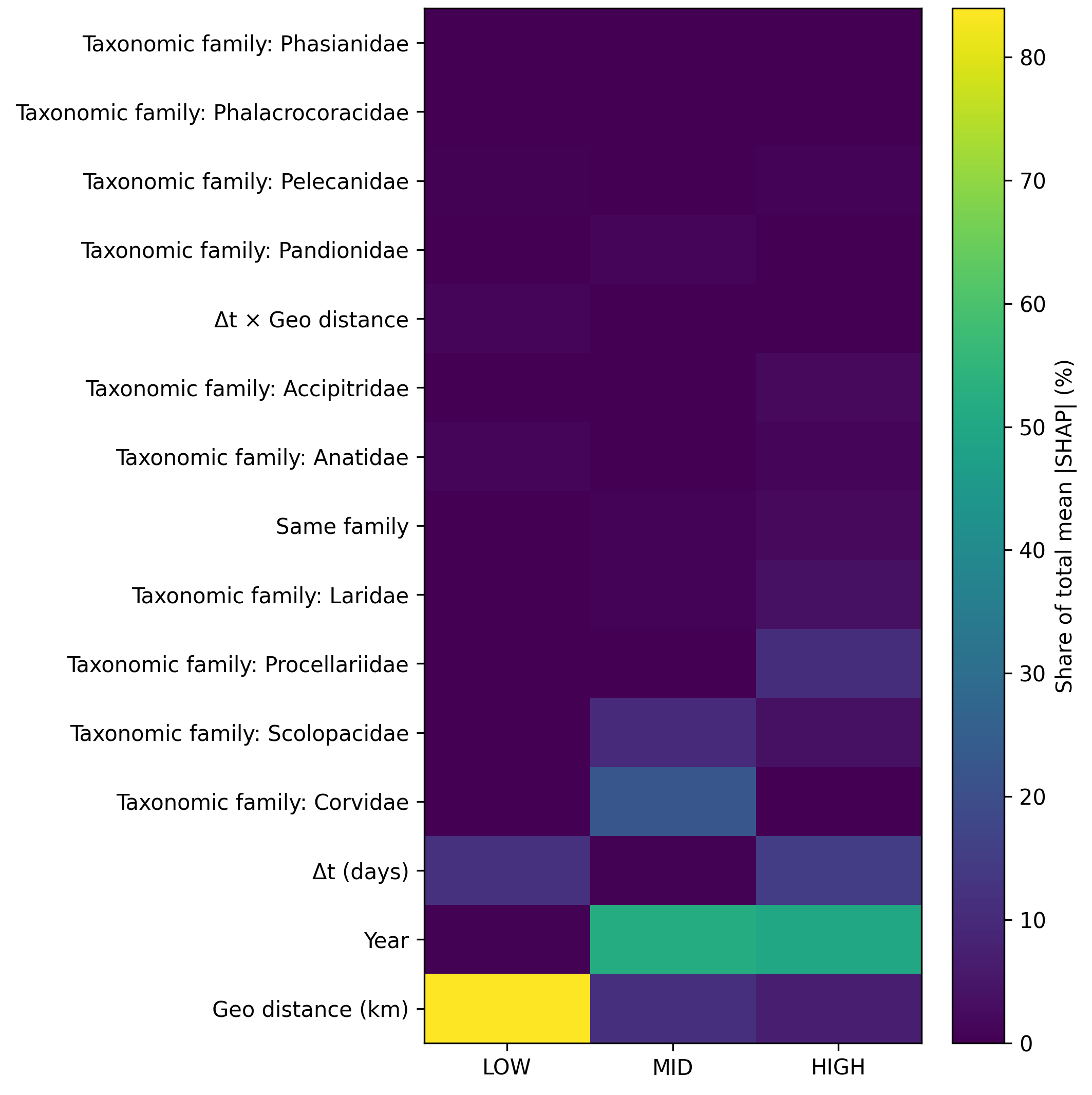}
\caption{Normalized SHAP importance for the median model ($\tau{=}0.5$) across divergence strata.}
\label{fig:shap-heatmap}
\end{figure}

We computed SHAP values for the median ($\tau{=}0.5$) models and report normalized importances (share of total mean $|\mathrm{SHAP}|$) across divergence strata in Fig.~\ref{fig:shap-heatmap}. In the LOW stratum, geographic distance is the dominant driver, with temporal lag contributing secondarily, indicating strong spatial–temporal coupling when evolutionary distances are small. In MID, the relative influence of year increases alongside geographic distance and host-related indicators, consistent with gradual temporal drift and changing host composition within circulating lineages. In HIGH, year and host-related indicators account for a larger share of importance, while spatial/temporal predictors contribute less—aligning with scenarios where larger genetic gaps arise from independent introductions or lineage structure rather than continuous diffusion. These patterns mirror the ablation study (Table~\ref{tab:ablation}) and help explain why point accuracy declines in HIGH while interval calibration remains stable.

\begin{figure*}
    \centering
\includegraphics[width=\linewidth]{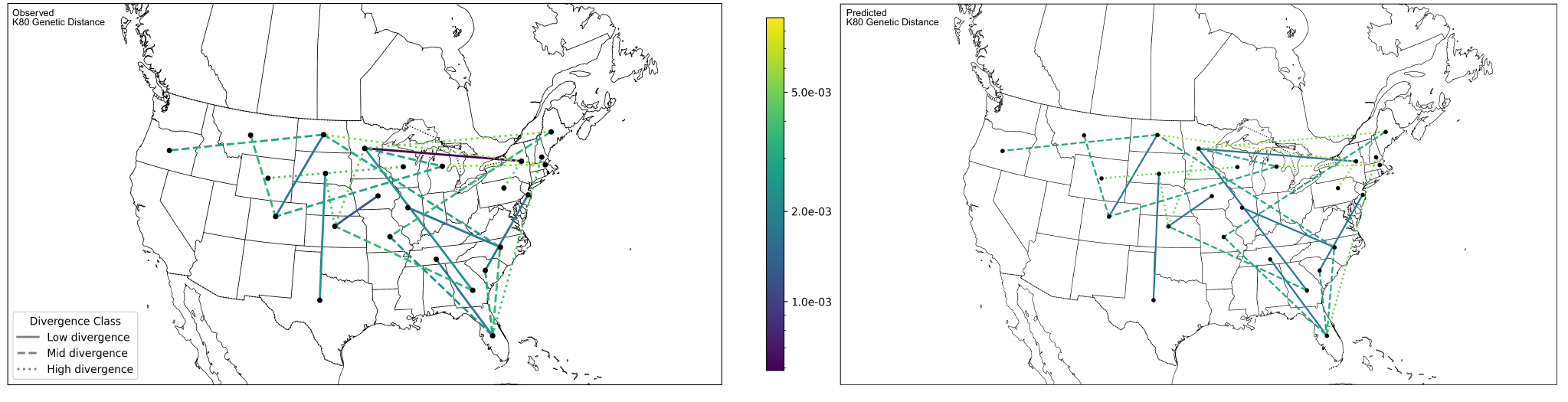}
    \caption{Spatial links between 30 A/H5 2022 case pairs in the United States, colored by observed (left) and predicted (right) K80 genetic distance.}
    \label{fig:casestudy}
    \vspace{-5mm}
\end{figure*}
\subsection{Case Study}

To illustrate model behavior across ecological and spatiotemporal contexts, we examined 30 A/H5 case pairs spanning LOW, MID, and HIGH divergence classes (Figure~\ref{fig:casestudy}). Each pair varied in collection date, geographic separation, and host taxonomy, allowing evaluation of how the QRI model responds to biologically relevant metadata.

LOW divergence pairs often involved closely related wild birds sampled in the same state or flyway. One example linked two snow goose detections in North Dakota, separated by 57 days and 0 km, with a true K80 distance of 0.0018. The predicted median was 0.0012 with a tight 90\% interval, consistent with minimal divergence under local transmission.

MID divergence examples typically included moderate geographic and host separation. A pair between a bald eagle in Minnesota and a wood duck in Michigan, collected 60 days apart and 787 km distant, had a true K80 of 0.0029. The model predicted a median of 0.0029 with well-aligned intervals, suggesting continuity across regional migratory paths.

HIGH divergence pairs reflected larger ecological or geographic gaps. For instance, a case pair between a duck in Wisconsin and a great horned owl in Wyoming, separated by 22 days and 1,433 km, exhibited a true K80 of 0.0041. The predicted interval was wide (0.0047 to 0.0069), reflecting elevated divergence consistent with cross-host or independent introductions.

Across all cases, the model adapted predictions to match spatial scale, host taxonomy, and temporal context. This enables biologically informed distance estimates where observed divergence is unavailable, and supports integration of unsequenced detections into evolutionary graph models.

\section{Discussion and Conclusion}
Imputing genetic divergence from metadata enables the integration of unsequenced cases into structured epidemiological analyses, addressing a key limitation in pathogen surveillance systems where genomic data are incomplete. The quantile regression framework presented here leverages temporal, spatial, and ecological covariates to estimate biologically plausible divergence intervals, without requiring sequence alignment or phylogenetic reconstruction. This approach captures meaningful evolutionary signal at low and moderate divergence levels, enabling high-resolution augmentation of diffusion models, cluster detection tools, and genomic surveillance platforms.

Although demonstrated on highly pathogenic avian influenza A/H5, the framework is pathogen-agnostic. Any setting with rich pairwise metadata including temporal lag, spatial separation, host or demographic attributes and observed divergence for a subset of pairs can be used. Adapting to other pathogens would require selecting an appropriate evolutionary substitution model and adjusting temporal scales to match evolutionary rates.

At higher divergence levels, predictive performance declines due to weaker correlations between the available metadata features and the true evolutionary distance. This effect is most pronounced in cases where divergence is driven by between-lineage differences, reassortment events, or other major genomic changes that are not explicitly represented in the model’s feature set. Without lineage or subclade identifiers, the model relies solely on spatiotemporal and host covariates, which have limited explanatory power in such contexts. These limitations are expected in settings characterised by ecological discontinuities, long-range spread, or independent introductions into distinct host populations, where genetic divergence may not follow patterns of continuous spatial diffusion. Consistent with this interpretation, the SHAP summaries for the median model (Fig.~\ref{fig:shap-heatmap}) show a shift in importance away from spatial and temporal predictors and toward year and host-related indicators in the HIGH stratum, mirroring the ablation results (Table~\ref{tab:ablation}). While this constrains accuracy in the HIGH divergence stratum, the model nonetheless produces well-calibrated uncertainty intervals across a broad range of surveillance-relevant cases, demonstrating robustness in realistic outbreak scenarios.

By shifting the focus from sequence-based inference to metadata-driven estimation, this framework provides a scalable, interpretable, and domain-informed solution for working with partially observed genetic datasets. Future work may extend this approach to incorporate lineage annotations, environmental drivers, or cross-segment dynamics, enabling richer models of viral evolution under data-sparse conditions. As global genomic surveillance expands, methods like this offer a pathway for integrating heterogeneous data streams, including spatial metadata, ecological context, and evolutionary information into unified forecasting and spatial decision support systems for epidemic response.

\section{Acknowledgements}
This research is supported by the Australian Commonwealth Scientific and Industrial Research
Organisation (CSIRO) and the the United States National Science Foundation under Grants No.
2302968, No. 2302969, and No. 2302970, titled "Collaborative Research: NSF-CSIRO: HCC: Small: Understanding Bias in AI Models for the Prediction of Infectious Disease Spread". This research is conducted by the ARC Centre of Excellence for Automated Decision-Making and Society (No. CE200100005) funded by the Australian Government through the Australian Research Council.We acknowledge the utilization of computational resources from the Katana High Performance Computing (HPC) cluster, which is supported by the Faculty of Engineering, UNSW Sydney.

\bibliographystyle{ACM-Reference-Format}
\bibliography{references}

\end{document}